\documentclass{article}
\usepackage{jheppubmod}
\usepackage{epsfig,amssymb,amsmath,latexsym}
\usepackage{hyperref}

\usepackage{color}
\definecolor{nicered}{rgb}{0.7,0.1,0.1}
\definecolor{nicegreen}{rgb}{0.1,0.5,0.1}

\newcommand{\be}{\begin{equation}}
\newcommand{\ee}{\end{equation}}
\newcommand{\bea}{\begin{eqnarray}}
\newcommand{\eea}{\end{eqnarray}}
\newcommand\ds{\displaystyle}
\newcommand{\no}{\noindent}
\newcommand{\nb}{\nonumber}

\newcommand{\de}{\partial}

\newcommand\E{{\mathcal{E}}}
\newcommand\K{{\mathcal{K}}}
\newcommand\F{{\cal F}}
\newcommand\Q{{\cal Q}}
\renewcommand\H{{\cal H}}
\renewcommand\S{{\cal S}}
\newcommand\T{{\cal T}}

\newcommand\C{{\cal C}}

\renewcommand\l{\lambda}

\renewcommand\a{\alpha}
\newcommand\D{\text{DoF}}

\renewcommand\k{\kappa}

\newcommand\m{\mu}
\newcommand\n{\nu}
\newcommand\g{\gamma}

\newcommand\V{{\ensuremath{\cal V}}}
\newcommand\tr{\text{Tr}}

\renewcommand\l{\ensuremath{\lambda}}

\renewcommand\L{\ensuremath{{\cal L}}}

\newcommand\ba{\begin{array}}
\newcommand\ea{\end{array}}

\newcommand{\plm}{M_{\text{PL}}^2}
\newcommand{\dd}{\delta^{(3)}(x-y)}

\newcommand{\U}{{\cal U}}

\title{Massive gravity: a General Analysis} 
\date{\today}
\author[a]{D. Comelli,}
\author[b]{F. Nesti,}
\author[c,d]{and L. Pilo}
\affiliation[a]{INFN, Sezione di Ferrara,  I-35131 Ferrara, Italy}
\affiliation[b]{Gran Sasso Science Institute, viale Crispi 7, I-67100 L'Aquila, Italy}
\affiliation[c]{Dipartimento di Scienze Fisiche e Chimiche, Universit\`a di L'Aquila,  I-67010 L'Aquila, Italy}
\affiliation[d]{INFN, Laboratori Nazionali del Gran Sasso, I-67010 Assergi, Italy}
\emailAdd{comelli@fe.infn.it}
\emailAdd{nesti@aquila.infn.it}
\emailAdd{luigi.pilo@aquila.infn.it}

\date{\small \today}

\abstract{Massive gravity can be described by adding to the Einstein-Hilbert action a function $V$
  of metric components.  By using the Hamiltonian canonical analysis, we find the most general form
  of $V$ such that five degrees of freedom propagate non perturbatively.  The construction is based
  on a set of differential equations for $V$, that remarkably can be solved in terms of two
  arbitrary functions.  Besides recovering the known ``Lorentz invariant'' massive gravity theory,
  we find an entirely new class of solutions, with healthy features on the phenomenological side, in
  particular they are weakly coupled in the solar system and have a high ultraviolet cutoff
  $\Lambda_2=(m M_{pl})^{1/2}$, where $m$ is the graviton mass scale.}

\begin{document}
\maketitle
\section{Introduction}
Whether general relativity (GR) is an isolated theory is an interesting question both from the
theoretical and phenomenological side.  In particular, the long standing quest for a healthy
modification of gravity at large distances recently has been target of a renewed interest.  The
attempt to find non-derivative, namely massive, deformations of GR dates back to the work of Fierz
and Pauli~\cite{Fierz:1939ix} (FP), where it was realized that, around Minkowski space, a generic
Lorentz-invariant (LI) massive deformation of gravity propagates 6 degrees of freedom (DoF) already
at quadratic level, and that the sixth mode is a ghost.  While one can tune the mass term to get rid
of the sixth mode at quadratic level, it reappears at nonlinear level or around non-flat
backgrounds~\cite{BD}.  Surprisingly, for a long time, a complete canonical study to understand the
propagating degrees of freedom non perturbatively, in a generic deformed version of GR was
missing. The gap was filled in~\cite{uscan} where it was shown that five DoF propagate if the
deforming potential solves a Monge-Ampere supplemented with an extra
equation~\footnote{See~\cite{Soloviev:2013mia} for a alternative analysis using Kuchar's Hamiltonian
  formalism.}. As matter of fact, infinitely many solutions exist, and the one discovered
recently~\cite{Gabadadze:2011,GF,DGT}~\footnote{This turned out to be actually a rediscovery,
  because B. Zumino already in 1970 showed that a Lorentz invariant ghost free theory
  exists~\cite{Zumino:1970tu}.} is just one of them.  The peculiarity of this theory is that it
exhibits an accidental Lorentz symmetry around a single background, usually taken to be the flat
one.  Conversely, most of the massive gravity theories with five DoFs are intrinsically Lorentz
violating.  As discussed in section~\ref{equiv}, this possible residual Lorentz symmetry acts just
in the pure gravity sector, and is not the same symmetry invoked by the Einstein equivalence
principle, which is phenomenologically required. In fact, it is actually the opposite, this symmetry
is rather costly from a phenomenological point of view: the Lorentz-invariant theory at quadratic
level reduces to FP and it shares with FP the failure to reproduce the correct light bending, in
sharp contrast with GR.  The issue does not disappear in the limit of vanishing graviton mass, the
so called vDVZ discontinuity~\cite{DIS}.  A possible way out was proposed by Vainshtein
in~\cite{Vainshtein}, by arguing that the linear approximation cannot be applied in the solar system
and nonlinear effects restore the correct GR behaviour at short distances from a source, or in the
limit of vanishing graviton mass.  While this mechanism has been verified in a number of
models~\cite{vain2}, it is important to point out that for it to work, the theory has to rely on
strong nonlinearities even at the macroscopic scales of the solar system where the values of the
gravitational potential is small. Quantum corrections also seem to be important at macroscopic
scales~\cite{padilla}. This makes the theory essentially hard to handle, since one can not use
perturbation theory even in a two body problem like the motion of a planet around the Sun.  Finally,
another issue is the existence of acausal superluminal modes~\cite{Deser}.

As we are going to show, this problematic situation is avoided if Lorentz invariance in the gravitational sector
is not imposed.  As this concerns only the gravitational sector and the equivalence principle still
can be preserved, this scenario is not only phenomenologically allowed, but may also be considered
interesting.  At the quadratic level, a class of such theories that are free from the vDVZ
discontinuity are available~\cite{Rubakov,dub,PRLus,gaba,diego}.  Remarkably many of the solutions
of the Monge-Ampere problem are precisely of this type.

In this paper we give the construction based on the canonical analysis that leads to the general
massive gravity theories with five propagating DoFs.  It turns out that the deforming potential has
to satisfy a set of differential equations, whose general solutions can be found in terms of two
arbitrary functions.
The result is also important from a phenomenological point of view; indeed, it provides a set of new
theories~\cite{usweak} which are weakly coupled in the solar system and with a cutoff $\Lambda_2=(m
M_{pl})^{1/2}$, the largest possible in the absence of a fundamental Higgs mechanism for
gravity. Since Lorentz is broken directly in the background, the existence of possibly superluminal
modes does not immediately translates into acausality (there is just a preferred frame) and on the
contrary it may constitute a peculiar experimental signature.

The outline of the paper is the following: in section~\ref{equiv} we discuss the various
formulations of massive gravity, stressing the important role of the residual symmetries of the
deforming potential $V$.  In section~\ref{can} a detailed study of the number of propagating DoFs is
given for a generic nonderivative modification of GR encoded in $V$; the result is that five DoF are
present if $V$ satisfies the Monge-Ampere equation supplemented by an extra differential equation in
field space.  In section~\ref{sol} we find the general solution to these equations, which leads to
the most general potential $V$ which propagates five DoFs.  In section~\ref{expsol} the potential
$V$ is constructed explicitly in a number of cases.  In section~\ref{phen} the perturbation theory
around Minkowski space is considered.  Finally, in section~\ref{conc} we summarize our conclusions.

\section{Massive GR: From Lorentz Invariance to Rotational Invariance}

\label{equiv}
To deform GR into a massive theory we need to add nonderivative functions of the metric field
$g_{\m\n}$.  Because any scalar diff-invariant function of the metric is trivial, this can only be
done by breaking diffs and thus reducing the invariance properties of the action.  The best one can
do is to preserve a residual set of symmetries in some specific background(s).  Indeed, formally
this can be described through the addition of fixed, external, tensor quantities, with different
kind of resulting theories.\footnote{General covariance can be also left unbroken by the use of
  appropriate St\"uckelberg fields. In section~\ref{sec:stuck} we will discuss in detail this
  approach, which does not change the result of this section and of our work.}

\begin{itemize}
\item A traditional way was to add an external second metric $\tilde g_{\m\n}$, so that combining
  $g$ and $\tilde g$ one can build a $(1,1)$ tensor
\be
\label{eq:XLI}
X^\mu_\nu = g^{\mu \alpha} {\tilde g}_{\alpha \nu} 
\ee
that constitutes the basis for all the necessary non trivial scalars~\cite{DAM1}
\be
\tau_n = \tr(X^n) = \left(X^n \right)^\mu_\mu \, , \qquad n = 1,2,3,4\, .
\label{eq:taun}
\ee

Around the background $g_{\m\n}=\tilde g_{\m\n}$, the combinations~(\ref{eq:taun}) are clearly
invariant under the (global) Lorentz transformations defined by $\tilde g_{\m\n}$.  Because of this
residual symmetry, the theories built from these invariants were named
``Lorentz-invariant''.\footnote{One could also give a dynamical character to the extra metric adding
  its Einstein-Hilbert action, see for instance~\cite{DAM1,PRLus,myproc,spher,Volkov,energy,spher1},
  leading to a bimetric theory.  In the bimetric formulation there is no absolute object but clearly
  the dynamics is more complicated; for instance, at the linearized level there are two gravitons,
  one massive and one massless, and the interaction terms break the ``relative'' diffs acting on the
  two metrics.}

For instance, with the popular choice $\tilde g_{\m\n}=\eta_{\m\n}$ and by expanding $g_{\mu \nu} =
\eta_{\mu \nu} + h_{\mu \nu}$ one has
\be
\begin{split}
&a \, \left(\tau_1 -4 \right)^2 + b \, \left(\tau_2 - 2 \tau_1 + 4 \right)=   \left (a \,
  h_{\mu \nu} h^{\mu \nu} + b \, h^2 \right)  + O(h^3) \, ,
\end{split}
\ee
i.e.\ the Lorentz-invariant generalized Fierz-Pauli mass term.

\end{itemize}
 
We wish to stress that, due to the presence of the fixed external metric $\tilde g_{\m\n}$,
Lorentz-invariance of these theories is only present around the particular background
$g_{\m\n}=\tilde g_{\m\n}$. In any other background (for instance cosmological) this residual
Lorentz invariance is not present.  A breaking will appear through the interactions
in~(\ref{eq:taun}).  In fact, the residual Lorentz symmetry should be not confused with the same
symmetry entering in the formulation of the Einstein equivalence principle.  Indeed, in general
backgrounds it is not possible to find a locally free fall reference frame such that both $g_{\mu
  \nu}$ and $\tilde g_{\mu \nu}$ coincide with the Minkowski metric.  The best one can do is to
choose a suitable tetrad basis $e^a_\mu$ such that
\be
g_{ab} = \text{Diag}(-1, \, 1 , \, 1 , \, 1)=\eta_{ab} \, , 
\qquad \tilde g_{ab} =  \text{Diag} (\lambda_0, \,
 \lambda_1 , \, \lambda_2 , \, \lambda_3) \, .
\ee
Thus, in general, local Lorentz invariance is never present in generic backgrounds, even in the so
called ``Lorentz-invariant'' theory.

Still, if the matter sector is minimally coupled only to $g_{\mu \nu}$, in this sector local
Lorentz invariance is always present, regardless of the invariance properties of the gravitational
sector.  No Lorentz-breaking or violations of the weak-equivalence principle will be observable from
the matter sector, and any breaking will be present only in the gravitational one, i.e.\ will
manifest itself in propagation of gravitational waves, etc.

In view of this, considering that there are so far no constraints on the Lorentz-invariance of the
pure gravitational sector, it makes sense to consider also the possibility of a smaller residual
symmetry.

\begin{itemize}

\item An immediate simple option is that of a Galilean SO(3) structure, which is achieved by
  introducing in the Lagrangian rotational invariants only.  In practice one can build SO(3)
  invariants using the metric components $g_{00}$, $g_{0i}$ and $g_{ij}$ (choosing for instance a
 canonical time-slicing $x^0=const$, but see later for comments on dependence on the slicing) or
  also better using the Arnowitt-Deser-Misner (ADM)~\cite{ADM} variables: lapse $N$, shift $N^i$,
  and spatial 3-metric $\g_{ij}$ ($i,j$=1,2,3),
  \be
  g_{\mu \nu} = \begin{pmatrix} - N^2 + N^i N^j \g_{ij} & \g_{ij}N^j \\
    \g_{ij}N^j & \g_{ij} \end{pmatrix} , \qquad g^{\mu \nu}
  = \begin{pmatrix} - N^{-2}  & N^{-2} \, N^i \\
    N^{-2} \, N^i & \g^{ij} - N^{-2} N^i N^j \end{pmatrix} .
\label{ADM}
\ee 

Again, to give mass to the graviton one needs to break spatial diffeomorphisms, so that an external
spatial metric has to be introduced.  For instance, one can use the euclidean cartesian metric
$\delta_{ij}$.  

As before, unbroken rotations will be a feature of the theory only around the background
$g_{ij}\propto \delta_{ij}$.  This is however sufficiently wide to include the homogeneous CMB
reference background, any homogeneous representation of the FRW cosmologies, etc.

Clearly, this approach includes the above Lorentz-invariant as a subcase, for one can decompose the
$\tau_n$ under rotations, e.g.\ $\tau_1 = -g^{00}+ g^{ij} \delta_{ij}$ (for $\tilde
g_{\m\n}=\eta_{\m\n}$, while in general one has also external fixed ``shifts'' in the game). It
should also be clear that a generic rotationally invariant function of the ADM variables can be
Lorentz invariant if and only if the variables appear in the particular combinations $\tau_n$.

\end{itemize}

Summing up, a general action of massive modified gravity theory can be written as
\be
S =  \int d^4x \sqrt{g}  \Big [M_{pl}^2  ( R 
  - m^2 \, V(N,N^i,\g_{ij})  + L_{\text{matter}} \Big]\, .
\label{act}
\ee
Matter is described by $ L_{\text{matter}}$ and is minimally coupled to the spacetime metric $g_{\mu
  \nu}$, in agreement with the equivalence principle.  The potential $V$, taken as a scalar function
of the ADM variables plus a fixed spatial metric, parametrizes the most general rotationally
invariant theory of massive gravity, and boils down to the Lorentz-invariant case if $V$ is actually
a function of the $\tau_n$.

The analysis in this work will be concerned with determining the form of $V$ to have a given number
of DoF at nonperturbative level.  As noted in~\cite{usweak}, the possible Lorentz breaking nature of
the resulting potentials opens a new window in the space of modified gravity theories to overcome
the critical drawbacks of the Lorentz-invariant case (most notably the dramatic strongly-coupled
nature of the theory at phenomenologically interesting scales). The necessity of breaking Lorentz
symmetry in the gravitational sector to have a healthy large-distance modification of gravity should
be considered as an interesting prediction to be experimentally tested.

It is worth stressing that with the non-dynamical metrics, which are effectively aether-like
absolute objects, preferred frame effects must be expected. Indeed, the frame where the frozen
metric assumes the form $\eta_{ab}$ is special and must be specified in relation with matter frame.
For instance, one could take such a frame as the one where the universe is homogeneous at large
scale (the CMB frame); this choice is particularly natural if massive gravity has something to do
with dark energy. Once such a physical choice is made, one should compare against standard preferred
frame constraints at the post-Newtonian level~\cite{will}. In the existing literature those effects
apparently have been over looked.

\section{Canonical Analysis}
\label{can}
Let us review the canonical analysis for a generic massive gravity theory given in~\cite{uscan}.  We
use the ADM Hamiltonian formulation in terms of lapse $N$, the shift $N^i$ and the spatial 3-metric
$\g_{ij}$.
The presence of the deforming potential violates diff invariance and thus the canonical analysis
drastically differs from the one in GR.  Following the very same steps of the ADM analysis, the
Hamiltonian relative to the gravitational part of action (\ref{act}) is
\be
H= \plm \int d^3x \left[N^A \,  \H_A + \, m^2 \,  N  \sqrt{\g} \, V  \right] ,
\label{ham}
\ee
where we introduced the notation $N^A=N_A=(N, \,N^i)$ and $\H_A = (\H, \,\H_i)$, with $A=0,1,2,3$.
The phase space is a 20 dimensional space of the conjugate variables ($\g_{ij}$, $\Pi^{ij}$) and
($N^A, \Pi_A$).  We will denote by $F_{,A}$ the derivative $\partial F/\partial N^A$ of any function
$F$, and we define $\V= m^2 \, N \, \sqrt{\g} \,V $.  As matter of fact, $\H_A$ are functions of the
spatial metric $\g_{ij}$ and its conjugate momenta $\Pi^{ij}$ only. Their (standard) form is given
for completeness in appendix~\ref{comm}.

When $m=0$, the action (\ref{act}) reduces to the familiar ADM GR Hamiltonian derived from the
Einstein-Hilbert action. In general no residual symmetry of potential $V$ will be assumed, thus our
analysis applies both to the Lorentz invariant (LI) case as to any nonlinear version of Lorentz
breaking (LB) models considered in~\cite{Rubakov,dub}.

Peculiar of GR and of also the deformed version (\ref{ham}), is that $N^A$ are non dynamical and
their conjugate momenta $\Pi_A$ are vanishing on the physical phase space (the $\Pi_A$ are called
\emph{primary constraints}).  The total Hamiltonian is
\be
H_T = H + \int d^3x \,  \lambda^A \Pi_A \, ,
\ee
where a set of Lagrange multipliers $\lambda^A$ are introduced to enforce the primary constraints
$\Pi_A \approx 0$.\footnote{The customary notation $\Pi_A \approx 0$ means that $\Pi_A$ are only
  weakly zero, e.g. zero on the constraints surface only.}  
  The time evolution of any function defined on the phase space 
   is given by:
\be
\begin{split}
\frac{d f(t,\vec x)}{dt} =&\, \{ f(t,\vec x), H_T(t) \} = \{ f(t,\vec x), H(t)\}+
 \int d^3 y \, \lambda^A(t,\vec y)
\{F(t,\vec x), \Pi_A(t,\vec y) \} \, .
\end{split}
\ee
The Poisson bracket between two functionals $F$ and $G$ of canonical variables $q_a,p^a$ is
defined as
$$
\{F, \,G \} 
=\! \int \!d^3 y \left [ \frac{\delta F}{\delta q_a(t,\vec y)} \frac{\delta
  G}{\delta p^a(t,\vec y)} -  \frac{\delta F}{\delta p_a(t,\vec y)} \frac{\delta
  G}{\delta q^a(t, \vec y)} \right] .
$$
In the following the dependence on space and time of the various fields will be understood.

Before studying the equations of motion of the dynamical variables $h_{ij}$, it is crucial to
enforce that the four primary constraints are conserved in time. This leads to \emph{secondary
  constraints} $\S_A$:
\be
 \S_A\equiv\{\Pi_A, H_T \} =- (\H_A + \V_{,A}) \approx 0 \, .
\label{second}
\ee 
Basically, these are the equations of motion of $N^A$. They do not determine any of the Lagrange
multipliers and are not automatically satisfied. Therefore, $\S_A\approx 0$ reduce the dimension of
the physical phase space by 4. Also $\S_A$ must be conserved, leading to new conditions, the
\emph{``tertiary'' constraints}
\bea
\label{ter}
&&\T_A \equiv \{ \S_A, \, H_T \} = \{ \S_A, \, H \}
- \V_{AB} \, \lambda^B \approx 0 \, ,\quad {\rm with}
\qquad \V_{AB}\equiv \V_{,AB}=\frac{\de^2 \V}{\de N^A \de N^B}\, . 
\eea
If the Hessian $|\V_{AB}|$ is non-degenerate, it can be inverted and the four eqs.~(\ref{ter})\ can
be solved for all the $\lambda^A$; all constraints are consistent with time evolution and the
procedure stops.

Summarizing, in the general case for which $\det|\V_{AB}| \neq 0$, we have $4\, (\Pi_A)+ 4\,
(\S_A)=8$ constraints, for a total of $(20-8)/2=6$ propagating DoF.  Technically, these 8
constraints are second class, being Rank$|\{ \Pi_A,\,\S_B\}|=$ Rank$|\V_{AB}|=4$. As a result, the
whole gauge invariances of GR are broken.

In the LI case Lorentz invariance tells us that the propagating DoF must be grouped in a massive
spin two (5 DoF) representation plus a scalar (1 DoF).  This is the Boulware-Deser result, valid for
a generic potential.  The extra scalar, the so called Boulware-Deser sixth mode~\cite{BD}, turns out
to be a ghost rendering these generic nonlinear theories unviable.

Thus, the first important result is that a necessary condition to have less than six propagating
DoF, is that $r \equiv\text{Rank}|\V_{AB}|<4$.

\subsection{The case $r=3$: five DoF} 

In this important case the matrix $\V_{AB}$ has a null eigenvector $\chi^A$ and three other
eigenvectors $E_n^A$ $(n=1,2,3)$ with eigenvalues $\k_n$:
\be
\V_{AB} \, \chi^B = 0\,,\qquad\V_{AB} \, E_n^B = \k_n \, E_n^A\, .
\label{null}
\ee
For instance if $\det(\V_{ij}) \neq 0$, then $\chi^A=(1, - \V^{-1}_{ij} \, \V_{0
  j})$.  The Lagrange multipliers can also be split along $\chi^A$ and its orthogonal complement
\be
\lambda^A = z \, \chi^A + \sum_{n=1}^{3}\, d_n \,E^A_n \stackrel{\text{def}}{\equiv}  z \, \chi^A +{\bar
  \lambda}^A\, .
\label{decomp}
\ee 
The Hessian is now non-degenerate only in the subspace orthogonal to $\chi^A$, and from (\ref{ter})\
one can determine only three out of four Lagrange multipliers, $d_n = E^A_n\{ \S_A, \,H \}/\k_n$.
The projection along $\chi^A$ leaves $z$ undetermined and~(\ref{ter}) gives a new non-trivial
(tertiary) constraint
\be
\T \equiv \chi^A \, \{\S_A, H \} \approx 0 \, . 
\ee  
Here and in the following we suppose that the constraints found are non trivial. At this point,
either the conservation of $\T $ allows to determine the remaining multiplier $z$ or it may generate
a further constraint. One has the \emph{``quaternary'' constraint}
\bea
\label{quat}
&\Q = \{ \T, H_T \} =  \{ \T, H \} + \{ \T,
\lambda^A \cdot \Pi_A \} \, , \quad {\rm with}\quad 
\lambda^A \cdot \Pi_A \equiv \int d^3y \, \lambda^A(y) \Pi_A(y) \, .
\eea
Clearly, the first piece of (\ref{quat}) does not contain $z$. Using (\ref{null})\ and
(\ref{decomp}) after some computation (see \ref{cantime} for more
details)  we find for the second piece
\bea\nonumber
  \{ \T,
\lambda^A \cdot \Pi_A \} &\approx&  \chi^B \{ {\bar
  \lambda}^A \cdot \S_A, \S_B  \} -  \chi^B \{  {\bar \lambda}^A \V_{BA} , H \} 
+\frac{\de \chi^B}{\de N^A}\, {\bar \lambda}^A \,{\bar \lambda}^D\, \V_{BD} - \Theta \cdot  z \,,
\eea
\vspace*{-2ex}
\be
\Theta(x,y) =\chi^A(x)\,  \{S_A(x),S_B(y) \} \, \chi^B(y) \, .
\ee
It turns out that, see appendix~\ref{comm},  $\Theta(x,y) = A^i(x,y) \de_i \delta^{(3)}(x-y) $ with $A(x,y)=A(y,x)$; it is
worth to point out that in a finite dimensional context $A^i$ would be absent. Thus,
\be
\Theta \cdot z = \int \!d^3 y \,\, \Theta(x,y) z(y) 
=  - \frac{1}{2z(x)}\de_i \left[ z(x)^2 A^i(x,x) \right].
\ee
As a result, $\Q$ is free from $z$ if $A^i(x,x)=0$, which consists in the following condition
\be
{\chi^0}^2 \,  \,\tilde  \V_i  + 2  \, \chi^A  \chi^j
\,\frac{\de \tilde  \V_A}{\de \gamma^{ij}} \,
 =0  \, , \qquad \V= \gamma^{1/2}   \tilde \V \, .
\label{add}
\ee
If this condition is satisfied, $\Q$ in (\ref{quat})\ is a quaternary constraint, which together
with $\T$ confirms the elimination of one  propagating degrees of
freedom (DoF).  If the condition (\ref{add}) is not satisfied,
(\ref{quat}) is not a constraint but can be used to determine the last Lagrange multiplier $z$. This
leaves us with an odd number of second class constraints and an odd dimensional physical phase
space, in particular $5+1/2$ DoF.  A similar situation is encountered in other deformations of GR,
see for instance~\cite{Henneaux:2010vx}.  It not clear if such a peculiar occurrence is physically
acceptable or not. For sure, condition (\ref{add}) is rather
restrictive.

When (\ref{add}) is satisfied by $\V$, the time evolution of the new constraint $\Q$ is non
trivial and can finally determine the multiplier
\be
z = -\left(\chi^A \,\Q_{,A} \right)^{-1} \left[ \,\bar \lambda^B \,\Q_{,B} +\{\Q, H \} \right]\,.
\ee
In conclusion, for $r=3$, there are $4\, (\Pi_A)+ 4 \, (\S_A) + 1\, (\T)+1 \, (\Q)=10$ constraints,
giving $ (20-10)/2 =5$ DoF, that is a good candidate for a theory of a massive spin-2.  The
required conditions are ({\ref{add})\ and
\be
{\rm Rank}|\V_{AB}|=3\;\;\;\  {\rm and}\;\;\; \ \chi^A \, \Q_{,A} \neq 0\,.
\ee
To our knowledge, the observation that a singular Hessian is a necessary condition to avoid the
Boulware-Deser argument which leads to six propagating DoF was made for the first time in general
terms in~\cite{DGT}. We would like to note that, though it is not straightforward, the same
conclusion is reached if we allow for an explicit time dependence in the potential $V$.

\subsection{General case} Let us briefly discuss the generic case in which the Hessian has any rank
$r$ between zero and four.  In this case $\V_{AB}$ has $4-r$ null eigenvectors $\chi^A_\alpha$,
\be
\V_{AB}\,\chi^B_\alpha=0\,,\qquad \alpha=1,\dots,4-r\,,
\ee
and as before, we can split the Lagrange multipliers as
\be
\lambda^A=\sum_{\alpha=1}^{4-r}\,z_\alpha\,
\chi_{\alpha}^A+\sum_{n=1}^{r}\, d_n \, E^A_n \, .
\ee
The projections of (\ref{ter})\ along $E^A_n$ determine the $r$ Lagrange multipliers $d_n=
E^A_n\{\S_A,H\}/\k_n$, while the projections along the $\chi_\alpha^A$ give $4-r$ tertiary
constraints
\be
\T_\alpha \equiv \chi_\alpha^A\,\{\S_A,H\}\approx 0\,.
\ee
The conservation of these constraints leads to
\be
\Q_\alpha= \{ \T_\alpha,H\} 
+\sum_{n=1}^r
\{ \T_\alpha , \Pi_A \} \cdot  d_n  E^A_n  
-\!\sum_{\beta=1}^{4-r} \theta_{\alpha\beta} \,\cdot\, z_\beta 
\label{fifth}
\ee
which are $4-r$ relations linear in the remaining $4-r$ Lagrange multipliers $z_\alpha$. The number
of DoF thus depends on how many multipliers are determined by these conditions.  The matrix
$\theta_{\alpha \beta}$ is 
\be
{\cal \theta}_{\alpha \beta}(x,y)
\equiv\;
\chi_{\alpha}^A(t,x) \,\{\S_A(t,x),\S_B(t,y)\}\,\chi_{\beta}^B(t,y) \, .
\ee
Notice that due to the dependence on $x$, $y$, the matrix $\theta_{\alpha \beta}$ is not necessarily
antisymmetric and its rank $s$ is not always even.
If $s<4-r$, some $z_\alpha$ are undetermined and we have $4-r-s$ new quartic constraints which
reduce the DoF. 

At the next (fifth) step, if the conservation of quartic constraints determines all the needed
multipliers, the procedure stops and we have only $16-2r-s$ constraints: $4\, (\Pi_A)+ 4 \,
(\S_A) +( 4-r)\, ({\cal T_\alpha})+(4-r-\,s) \, (\Q_\a)$. This implies that the maximal number of
DoF is
\be
 \D\le 4+r/2 \, .
\ee
This is the maximal number because if some of the $z_\a$ are not determined, further steps are
necessary possibly reducing the number of DoF.  In general also first class constraint may be
present corresponding to residual gauge invariances, reducing even more the number of DoF.

\medskip

Summarizing, massive deformations with 5 DoF exist only in two cases: $r=3$, $s=0$ and $r=2$,
$s=2$. In the case $r=3$, the condition $s=0$ is equivalent to the differential equation (\ref{add})
for the deforming potential.  In the case $r=2$ and $s=2$, (\ref{fifth}) determines the remaining two
Lagrange multipliers and all 10 constraints are consistent with the time evolution. Notice that it
is not possible to build a potential with $s=2$ without breaking spatial rotations.  Thus, we have
shown that $r=\text{Rank}(\V_{AB}) =3$, together with (\ref{add}) are necessary and sufficient
conditions for a rotational invariant massive gravity theory with 5 DoF (again we consider that none
of the constraints found is accidentally trivial, i.e.\ $0 \approx0$).

\section{Solution for five DoF: the Monge-Ampere problem}
\label{sol}
We have shown that the potentials that have 5 DoF at nonperturbative level are solutions of the
following system of coupled differential equations in the space of fields $(N^A,\,\g_{ij})$
\bea
\label{monge}
&&\det(\V_{AB})=0\,,\\
\label{pilo}
&&\tilde{\V}_i+2\,\xi^A\;\xi^j\; \frac{\partial \tilde\V_A}{\partial\g^{ij}}=0\,,
\eea
where we defined $\xi^A=\chi^A/\chi^0$ and  $\tilde \V= \gamma^{-1/2}  {\V}$.
Since spatial rotations are preserved,
$\det(\V_{ij})\neq0 $ and  the $r=3$ condition can be rewritten into the equivalent form
\be
{\V}_{00}-\V_{0i}(\V_{ij})^{-1}\V_{j0}=0 \, .
\ee
Physically, this means that once the equation for the shifts $N^i$ are solved and inserted back in
the potential, $\V$ will be linear in $N$.  Mathematically, (\ref{monge}) is an homogeneous
Monge-Ampere equation~\cite{MongeAmpere} on the space of the $N^A$ variables, while $\g_{ij}$ is a
spectator variable.  Particular solutions of this equation are known in a closed form and,
remarkably, also the general solution can be given in parametric form~\cite{fai}, which we will exploit below.
Thus, there exist a large class of potentials, taken as a function of lapse, shift and spatial
metric $\V(N,N^i,\g_{ij})$ that satisfies~(\ref{monge}).  Note also that if $\V$ satisfies the
Monge-Ampere equation also $\tilde \V$ does, so we can substitute $\V\to\tilde \V$ into
(\ref{monge}).  Let us describe the general parametric solution.

Our starting point is the existence of a null eigenvector $\chi^A$ such that $\chi^A\; \tilde\V_{AB}=0$.
Then it is crucial to change variables from $N^A$ to $\chi^A$, and study the properties of the
function
\be \label{uv}
{\bf U}(\chi^A)\equiv \tilde \V_A \, \chi^A \, .
\ee
We easily see that ${\bf U}$ is an homogeneous function of  weight one,  i.e.
\be
\frac{\partial {\bf U}}{\partial \chi^A} = \tilde\V_A\quad\to\quad {\bf U}=\chi^A\;\frac{\partial {\bf U}}{\partial \chi^A}
\quad\to\quad {\bf U}(\k\;\chi^A)=\k\;{\bf U}(\chi^A)\,.
\label{aux}
\ee
Therefore we can write 
\be
 {\bf U}(\chi^A)\equiv \chi^0\;{\U}(\xi^i),\qquad \xi^i\equiv \frac{\chi^i}{\chi^0}\,.
\ee
where we have introduced a generic function $\U(\xi^i)$.  Everything can now be expressed in terms
of $\U$ and relations (\ref{uv}), (\ref{aux}) become
\bea
\label{basic1}
 \tilde\V_0&=&{\U}-\frac{\partial{\U}}{\partial \xi^j}\;\xi^j,\qquad
 \tilde\V_i = \frac{\partial{\U}}{\partial \xi^i}\,,
\eea
which allow one to reconstruct $\V$ for any given $\U$.  Note, in what follows we will denote with
$\U_i$, $\U_{ij}$, etc., the corresponding derivatives with respect to the $\xi^i$ (while instead 
$\tilde\V_i=\partial \tilde\V/\partial N^i$).

These equations have to be integrable, i.e.\ the following conditions have to be imposed:
\be\label{vc}
  \tilde\V_{ij}= \tilde\V_{ji} \quad{\rm and }\quad  \tilde\V_{0i}= \tilde\V_{i0}\,.
\ee
The first of these equations boils down to
\be
\U_{ij}\left(\frac{\partial \xi^j}{\partial N}+\frac{\partial \xi^j}{\partial N^k}\xi^k\right)=0
\quad\to\quad  \left(\frac{\partial \xi^j}{\partial N}+\frac{\partial \xi^j}{\partial N^k}\xi^k\right)=0\quad {\rm if}\quad 
 \left|\U_{ij}\right|\neq 0\,,
\label{flow}
\ee
which are known equations describing a generalized hydrodynamic
flow. Equations (\ref{flow}) can be solved~\cite{fai} by
means of an implicit relation between the $N^i$ and $\xi^i$,
exploiting also the second equation~(\ref{vc}), which reads
\vspace*{-1ex}
\bea
\label{Qrel}
&& N^i=N\; \xi^i+\Q^i,
\qquad\Q^i(\xi^i,\gamma^{ij})=-\left(\U_{ij}\right)^{-1}\E_j
\eea 
where $\E(\xi^i,\g^{ij})$ is a new arbitrary function (with again $\E_i=\partial_{\xi^i}\E$), while
$\left(\U_{ij}\right)^{-1}$ is the matrix inverse of the hessian of $\U$.  This is a crucially
important change of variables, and should be thought as defining implicitly $\xi^i(N^i)$.

Integrability being assured, one can find a potential $\V$ by integrating~(\ref{basic1}).
Remarkably, the result of the integration can be performed explicitly (see Appendix~\ref{int}) and
the result for the potential is\footnote{Note: during the integration of~(\ref{basic1}), the choice
  of an integration constant $\C(\gamma^{ij})$ is required. This can be reabsorbed in $\E$ without
  changing the previous steps. With respect to~\cite{usweak}, we have here $\E=\C-\L$.}
\be
\tilde\V = N\;\U+\U_i\,\Q^i+\E\,.
\ee
or 
\be
\label{Vgeneral}
V = \U+N^{-1} \;\left(\U_i\,\Q^i+\E \right) \,.
\ee
This completes the general solution of the Monge-Ampere equation.  

Summarizing, the solutions can be parametrized with two arbitrary functions $\U$ and $\E$.
Concretely, the strategy to generate explicit potentials function of $(N,N^i,\g^{ij})$ is therefore
first to choose $\U$ and $\E$ which already fully determine the potential, and then to
solve~(\ref{Qrel}) for $\xi^i$ in terms of ($N^i$, $N$, $\gamma_{ij}$).

\medskip

As a final step, we have to impose condition (\ref{pilo}) in order to fully reduce to five DoF in
phase space.  This brings $\g^{ij}$ into the game. It is nice that, when expressed in terms of
${\U}$,~(\ref{pilo}) takes an elegant and compact form
\be
\frac{\de \U}{\de \xi^i}  + 2 \,\xi^j \, \frac{\de \U}{\de \gamma^{ij}} =0
\, ,
\label{five}
\ee 
whose solution is readily found as
\be\label{uk}
\U=\U(\K^{ij}),\qquad{\rm with}\qquad \K^{ij} =  \gamma^{ij} -\xi^i\xi^j \,.
\ee
This completes the solution of our problem, which is described in terms of the two functions
$\U(\K^{ij})$ and $\E(\xi^i,\g^{ij})$.

\medskip

Since we have assumed unbroken rotations, $\U$ must depend only on scalar combinations built with
$\K^{ij}$ and $f_{ij}$ (and not with $\gamma_{ij}$ because this would invalidate eq.~(\ref{five})).
So, the rotationally invariant solutions of eq.~(\ref{five}) are of the form $\U\left(\text{Tr}[(\K f)] \,
, \text{Tr}[(\K f)^2], \text{Tr}[(\K f)^3]\right)$.

For generic $\U$ and $\E$, the resulting potential (\ref{Vgeneral}) will in general be Lorentz
breaking.  It is in fact clear the reason why a full Lorentz symmetry is not natural: the basic
scalars that solve~(\ref{five}) are just {\it three}, while there are {\em four} $SO(3,1)$
invariants built out of $X^\mu_\nu=g^{\mu\alpha} \eta_{\alpha \nu}$, for instance its
eigenvalues. Thus, in building the potential $V$ from $\U$, the {\it four} $SO(3,1)$ scalars should
come out only with a very special choice of the function $\E$ entering~(\ref{Qrel}) (e.g.\
see~(\ref{eq:LLI}) below). This is in agreement with the discussion above, in section~\ref{equiv}.

\medskip

Before describing explicit examples we can also evaluate the hamiltonian $H$, after the use of the
(secondary) constraints (we neglect the surface term that is formally identical to the one of
GR):
\bea\nonumber
H_{constrained}&=&\int d^3x \;\left(\H_A\;N^A+\V \right) _{ \tiny S_A\approx 0} \approx
\int d^3x \;\left(-\V_A\;N^A+\V \right)_{\text{eq}.(\ref{basic1})}\\
&=&\nonumber
\int d^3x \;\g^{1/2}\;\left(
\tilde \V- (\U-\U_j\; \xi^j)\;N-
\U_i\;N^i \right)_{\text{eq}.(\ref{Qrel})} \nb\\
&=&\int d^3x \;\g^{1/2}\;\left(
\tilde \V- \U \;N-\U_i\Q^i\right) \nb\\
&=&\int d^3x \;\g^{1/2} \,\E\,.
\label{energy}
\eea
Thanks to this remarkably simple expression, one can have positive definite energy by choosing
appropriately $\E$.

\section{Explicit theories with five DoF}
\label{expsol}
Now that we the general parametrized solution to the problem in terms of the free functions $\U$ and $\E$,
let us look for some specific examples where the potentials can be given in a closed form.

\subsection{Lorentz breaking solvable examples}

\paragraph{Case 1.}
A simple but quite wide set of theories is found by looking for $\E$ such that $\Q^i\propto\xi^i$:
\be
\Q^i = \zeta(\gamma)  \, \xi^i 
\ee
for any arbitrary scalar function $\zeta$ of $\gamma^{ij}$, so that
\be
\xi^i  = \frac{N^i }{N+\zeta} \qquad \text{and}\qquad
\K^{ij}=\g^{ij}-\frac{N^iN^j}{(N+\zeta)^2} \, .
\label{gen}
\ee
This is realized by the following $\E$
\be
\E = -\left(\xi^j \U_j- \U \right) \, \zeta+\C \, ,
\ee
for any function $\C$ of $\gamma^{ij}$. From this we find the potential
\be
\label{genpot2}
\tilde \V = \left(N+\zeta\right)\,
\U(\K) + {\cal C}(\g) .
\ee

\pagebreak[3]

The minimal case of $\zeta=0$ was considered in~\cite{usweak} and it already has a very interesting
phenomenology. We noted there that the choice $\zeta=0$ together with $\C=0$ is undesirable as it
leads to strong coupling in the scalar and vector sector.

\paragraph{Case 2.}  An other solvable case is if $\Q^i$ is linear in $\xi^i$
\be
\label{QM}
\Q^i= M^i_j \;\xi^j \, 
\ee
with $M$ a matrix that can depend on $\gamma^{ik}f_{kj}$.  Thus, (\ref{Qrel}) can be inverted to
express $\xi$ in terms of the lapse and shifts 
\be 
\vec \xi=(N+M)^{-1}\vec N\,.
\ee
where it is here convenient to use the matricial forms $\vec N=\{N^i\}$, $\vec\xi=\{\xi^i\}$, as
well as $\U''=\{\U_{ij}\}$, $\vec\E'=\{\E_i\}$, etc.  For $\E$, eq. (\ref{QM}) implies that 
\be 
\vec\E'=-\U''\, M \vec\xi \, , 
\ee
whose integrability conditions $\E_{ij}=\E_{ji}$ are
\be
 \U^{''} \,  M - M^t \, \U^{''}  =0 \,.
\ee
If $M$ is not the identity (which case corresponds exactly to case 1 above) this condition can be
satisfied only if $\U''$ is independent of $\vec\xi$.  Thus $\U$ must be at most quadratic in
$\xi^i$, and because it must be a function of $\K^{ij}=\g^{ij}-\xi^i\xi^j$, the only allowed term is
$\text{Tr}(\K\cdot f)$. As a result we have
\be 
\begin{split}
&\U =\tilde C  -\frac{c_1}{2} \, \text{Tr}(\K f)= \tilde C-\frac{c_1}{2}
\, \text{Tr}(\g^{-1}f)+\frac{c_1}{2}\,\vec\xi^T\, f\, 
\vec\xi \, ;\\
& \E = \C -\frac{c_1}{2} \,  \vec\xi^T (f \cdot M) \vec\xi \, .
\end{split}
\ee
where $\C, \, \tilde C$ are functions of $\gamma$ and $c_1$ is a constant.  The
potential $\V$ can be derived for general $M$ 
along the lines of the previous examples and clearly gives an
expression which is quadratic in the shifts 
\be
V=\frac{c_1}{2}\left[
\vec N^T  (N+M)^{-1}  \left(f+ N^{-1} M \cdot f \right)  (N+M)^{-1} \vec N -\;\text{Tr}(\g^{-1} f)\right]
+\tilde C + N^{-1} \,  C \, .
\ee
The corresponding on-shell Hamiltonian  (\ref{energy}) reads
\be
H= \int d^3x\, \gamma^{1/2} \left[-\frac{c_1}{2} \, \vec N\cdot (N+M)^{-1}\cdot
  f\cdot M\cdot(N+M)^{-1}\cdot \vec N +  C \right].
\ee 

\subsection{Lorentz Invariant case}
\label{sec:LI}

Let us consider now the Lorentz invariant case. The change of variables (\ref{Qrel}) required for
the solution of the Monge Ampere equation,
\be
N^i = N\,\xi^i + \Q^i(\xi,\gamma) 
\ee
coincides with the change of variables used in~\cite{GF} to show that the Lorentz invariant
potential propagates five DoF, after relating the variables $n^i$ to $\xi^i$, via $n^i(\xi,\gamma) =
\big(D^{-1}\big)^i_j \xi^j$.  Therefore, this represents a nontrivial implicit relation between
$N^i$ and $\xi^i$.  In detail we have (we limit ourselves to $f_{\m\n}=\eta_{\m\n}$) %
\bea 
(D^2\big)^i_j &=&\big[1+\xi\cdot \K^{-1}\cdot\xi\big]\ (\K\cdot f)^i_j\, ;\qquad
\Q^i= \big(D^{-1}\cdot\xi\big)^i\,,\\
 {\rm with} &&\K= \gamma^{ij}-\xi^i\xi^j \quad \text{and}\quad \K^{-1}= \gamma_{ij}+\frac{\xi_i\xi_j}{1-\xi^2}\,,
\eea
where $\xi_i=\g_{ij}\,\xi^j$ and $\xi^2\equiv \xi^i\,\g_{ij}\,\xi^j$.
The two unknown functions $\U$ and $\E$ are 
\be
\U_{LI} = \text{Tr}\left[(\K \cdot f)^{1/2} \right] -3 \, , \qquad \E =
\left(1+\xi\cdot \K^{-1}\cdot\xi \right)^{1/2} =\frac{1}{\sqrt{1-\xi^2}}
\label{eq:LLI}
\ee
and computing $\U_i = - [\left(\K \cdot f \right)^{-1/2} \cdot f \, \xi]_i$ we have
\be
\Q^i\; \U_i = -\frac{\xi\cdot\K^{-1}\cdot \xi}{\left(1+\xi \cdot \K^{-1}\cdot \xi
 \right)^{1/2}} =-\frac{\xi^2}{\sqrt{1-\xi^2}} ,
\ee
so that the full potential turns out to be
\be
\tilde \V =
N\;\left(\text{Tr}\left[(\K \cdot f)^{1/2} \right] -3\right)+\sqrt{1-\xi^2}=
\sqrt{1-\xi^2}\;\left( 1 + N \, \text{Tr}(D) \right) - 3 \, N \, . 
\ee
(Note, the action is real only for $\xi^2<1$.)  Thus, also in the Lorentz invariant case $\U$ is a
function of $\K$ only, and therefore it satisfies automatically the nontrivial
condition~(\ref{five}) for the complete elimination of the sixth mode in phase space.

Going back to the original variables~\cite{GF}, we recover the Lorentz invariant expression in terms
of $X^\mu_\nu = g^{\mu \alpha} f_{\alpha \nu} $:
\be
V_{LI} = \text{Tr}(X^{1/2}) - 3 \,,\qquad X^\mu_\nu = g^{\mu \alpha}f_{\alpha \nu}  \, .
\ee
The on-shell Hamiltonian is
\be
H=
 \int d^3x \, \frac{\gamma^{1/2}}{\sqrt{1-\xi^2}}   .
\ee
and it is real and positive only for $ \xi^2<1$ (which includes the flat space configuration).
Notice that actually if $\xi^2=1$, $K^{ij}$ is not invertible and the change of variables from
$\xi^i$ to $N^i$ is ill defined.

\section{St\"uckelberg Dictionary}
\label{sec:stuck}

Even if the theories considered in this work feature a massive graviton and thus a direct violation
of general covariance, this invariance can be restored, or considered never broken, by the use of
St\"uckelberg fields. This can be achieved by promoting the external metric(s) or tensorial
quantities to true tensors, in such a way that the action is fully diff invariant.

For instance, as is well known, in the case of Lorentz invariance, one can promote the external
metric in~(\ref{eq:XLI}) to a true tensor by using four St\"uckelberg fields $\Phi^A$ ($A=1,\ldots,4$):
\be
{\tilde g}_{\mu \nu} = \de_\mu \phi^A \de_\nu \phi^B \ \eta_{AB} \, ,
\ee
after which all traces of $X^\m_\n$ are also diff invariants. 

To promote rotational scalars to full diff invariant quantities one can introduce a similar set of
St\"uckelberg scalar fields, starting with ``universal time'' field $\Phi$ which defines a spatial
slice through $\Phi=const.$ with unit normal
$n_{\mu}=\de_\mu\Phi/(-g^{\alpha\beta}\de_\alpha\Phi\de_\beta\Phi)^{1/2}$.  In the spatial
hypersurface $\Phi=const.$ one also considers the 3D metric $f_{ij}$ that is pulled back to
spacetime with three more St\"uckelberg scalars $\Phi^i$
\be
f_{\mu \nu} = f_{ij}\; \de_\mu \Phi^i \, \de_\nu\Phi^j  \, . 
\ee
Of course, the presence of a residual $SO(3)$ symmetry in the unitary gauge requires that
$f_{ij}=\delta_{ij}$. Note that in this procedure, the preferred time slicing has become dynamic
with the gauge field $\Phi$.  In the unitary gauge, $\bar x^0 =\Phi$ and $\bar x^i= \Phi^i$ we have
that
\be
\frac{\de \Phi^i}{\de \bar x^\mu} =  \delta^i_\mu \, , \qquad\frac{\de
  \Phi}{\de \bar x^\mu} =  \delta^0_\mu \, , 
\qquad f_{\mu \nu}= \delta_\mu^i \;\delta_\nu^j \;  \delta_{i j}   \, .
\label{LBUG}
\ee
In general, any $SO(3)$ scalar in the unitary gauge can be written out of the following 4D basic
scalar, vector and tensor quantities
\be
n =  (-g^{\alpha\beta}\de_\alpha\Phi\de_\beta\Phi)^{-1/2}  \, , \quad
n_{\mu}=\frac{\de_\mu\Phi}{(-g^{\alpha\beta}\de_\alpha\Phi\de_\beta\Phi)^{1/2}}
\, , \quad Y^\mu_\nu = g^{\mu \alpha} f_{\alpha \nu} \, .
\label{obj}
\ee 
I.e.\ the $\tau_n$ of (\ref{eq:taun}) can be decomposed in terms of new scalar combinations of them,
for instance we have $\tau_1 = n^{-2} + g^{\mu \nu} f_{\mu \nu}$.

After this dictionary, all the potentials discussed in the previous sections can be easily written
in a generic coordinate system simply by noting the the lapse $N$ should be replaced by $ n$, all
scalar functions of $\gamma^{ik}\delta_{kj}$ replaced by the same scalar functions of
$\Gamma^\mu_\nu$ defined as
\be
\Gamma^\mu_\nu= \left(g^{\mu \alpha} +
  n^\mu \, n^\alpha \right ) f_{\alpha \nu}=Y^\mu_\nu + n^\mu \, n^\alpha  f_{\alpha \nu} \, .
\ee
which plays the role of the pull-back of the 3D spatial metric.

The potential for the case 1 is given by
\bea
&& V_{\text{\tiny case 1}}=n^{-1}\left[ n + \zeta(\Gamma)
\right] \U(\tilde \K)+ n^{-1} \C(\Gamma) \, .
\eea
where $\U$ is now a scalar function of
\be
 \tilde \K^\mu_\nu = \left( \Gamma^{\mu \alpha} -
  \frac{n^2 \, n^\mu n^\alpha}{\left[n +
      \zeta(\Gamma)\right]^2} \right) \, f_{\alpha \nu} \, . 
\ee
The expression for $V$ in the case 2 is not particular illuminating and it can be determined along
the above lines.

Thanks to the presence of the St\"uckelberg fields, the actions maintain the invariance under
diffeomorphism. Once the unitary gauge is chosen, only a residual Lorentz or rotational invariance
is left.

\section{Perturbations around flat space}
\label{phen} 
 
From the weak field expansion of the action (\ref{act}), taking $g_{\mu \nu} = \eta_{\mu \nu} +
h_{\mu \nu}$, at the quadratic order in $h$ we have
\be
S= \int d^4 x M_{pl}^2 \left[ \E_{(2)} + \frac{m^2}{2} \left(
    m_0^2 \, h_{00}^2 + 2 m_1^2 \, h_{0i} h_{0i} -m_2^2 \, h_{ij} h_{ij}
    + m_3^2 \, h_{ii}^2 -2  m_4^2 \, h_{00} h_{ii} \right) \right] \, ;
\label{masses}
\ee
where $ \E_{(2)}$ is the standard Lagrangian for a massless spin 2 particle in Minkowski space,
stemming from the Einstein-Hilbert action expansion. The other terms represent the most general form
of the expansion of any potential with a $SO(3)$ residual symmetry. The physical consequences of a
such mass term were first discussed in~\cite{Rubakov}.  

Here, using (\ref{Vgeneral}) the masses can be computed explicitly by differentiating the potential.
As the first step, one has to find the conditions for the potential to admit the Minkowski
background $g_{\mu \nu} = \eta_{\mu \nu}$. These are
\be
\bar \U =0 \, , \qquad \qquad 
\bar \U' + \bar \E' - \frac{\bar \E}{2}=0 \, .
\label{flat}
\ee 
The bar indicates that all expressions are evaluated on Minkowski space and $\de \E / \de
\gamma^{ij} \equiv \E' \, \gamma_{ij}$, $\de \U / \de \K^{ij} \equiv \U' \, \gamma_{ij}$. We have
also used the fact that by rotational invariance on Minkowski space $\left.{\Q_i}{}
\right|_{\eta}=\left.{\U_i}{}\right|_{\eta}=\left.\xi^i\right|_{\eta}=0$.  From the general
expression of the potential (\ref{Vgeneral}) and the condition (\ref{flat}), we find the masses
\be
\begin{split}
m_0^2 &=\left.-\frac{m^2}{4} \frac{\de^2 ( \sqrt{g} \;V)}{\de N^2}\right |_{\eta} =0 \; ;\\
m_1^2 &=\left.  - \frac{m^2}{2}  \frac{\de^2 ( \sqrt{g}\; V)}{\de N^i \de
  N^i}\right |_{\eta} =\left. m^2 \,  \bar \U'\; \left(\frac{\de \xi^i}{\de N^i}
\right)\right |_{\eta}   \; ;\\
m_4^2 &=\left.  \frac{m^2}{2}  \delta_{ij} \frac{\de^2 ( \sqrt{g} \;V)}{\de \gamma^{ij} \de
  N}\right |_{\eta} =\frac{ 3 \, m^2}{2} \,  \bar \U' \; .\\
\end{split}
\label{massval}
\ee
The expressions for $m_2^2$ and $m_3^2$ are not particularly illuminating.  It is reassuring that
the condition $m_0^2=0$, to have only 5 DoF~\cite{BD} at the linearized level, is automatically
satisfied. Moreover, the following relation holds
\be
m_1^2 = \left.\frac{2}{3}   \;m_4^2 \; \left(\frac{\de \xi^i}{\de N^i}
\right)\right |_{\eta} \, .
\ee
In a Lorentz invariant theory where $m_1^2=m_4^2$ one should have that $\left.\left(\de \xi^i / \de
    N^i \right)\right |_{\eta} = 3 / 2$, as can be checked from section~\ref{sec:LI}. In the Lorentz
breaking case, a large class of potentials easily passes the solar system test and are weakly
coupled in all phenomenological interesting scales~\cite{usweak}.

Let us briefly review the linearized analysis of~\cite{Rubakov} (see also~\cite{dub}).  It is
convenient to expand the perturbation $h_{\mu \nu}$ in terms of $SO(3)$ scalars, vectors and
tensors, namely
\be
h_{00} = \psi \, , \qquad h_{0i} = u_i + \de_i  v \, , \qquad h_{ij}= \chi_{ij}
+\de_i a_j + \de_j a_i + \de_i \de_j \sigma + \delta_{ij} \, \tau  \; .
\ee
It is easy to see that the transverse and traceless tensor $\chi$ has always a standard kinetic term
and a mass $m_2^2$; thus the only requirement is $m_2^2 >0$.  In the scalar sector, when $m_0^2=0$,
$\psi$ is non-dynamical at appears linearly in the quadratic action.  Again, from (\ref{massval}),
when $m_1 \neq 0$ also $m_4 \neq 0$ and we can solve for $\sigma$. We are left with a single
propagating scalar. In the vector sector $u_i$ is non dynamical and can be integrated out, leaving a
single propagating vector $a_i$.  As a result, we have five DoFs that match canonical analysis. On
the other hand, if $m_1 =0$, in the vector sector there is no propagating DoF. But since from the
nonperturbative canonical analysis we already know that actually there are 5 DoFs, this means that
the vanishing of the kinetic term in the vector sector when $m_1=0$ it is just an accident of the
linearized expansion around Minkowski space. This signals that there is a vector mode that is
strongly coupled.

For a static source with a non-zero energy momentum tensor component $t_{00}$, the gauge invariant
potential $\Phi=h_{00}$ is~\cite{Berezhiani:2009kv}
\bea 
\label{eq:yuk}
\Phi=2\,t_{00}\frac{\Delta+m_2^2\,\frac{m_2^2-3\,
    m_3^2}{m_2^2-m_3^2}}{4\,\Delta^2+
2\,\Delta \,m_4^2\,\frac{m_4^2-4\,m_2^2}{m_2^2-m_3^2}+6\,m_4^4\, \frac{m_2^2}{m_2^2-m_3^2}}
\equiv\frac{t_{00}}{2}\left[\frac{A_1}{\Delta-M_1^2}+\frac{A_2}{\Delta-M_2^2}\right] 
\eea 
where the squared masses $M_{1,2}^2$ and the coefficients $A_{1,2}^2$ in the formal decomposition of
the  last term depend on the mass parameters $m_i^2$, with $A_1+A_2=1$.  It is straightforward to
check that under the mild conditions
\bea 
0<m_4^2<4 \,m^2_2\,, \qquad \ds-\frac{m^4_4}{24\; m^2_2}+\frac{m^2_4}{3}+\frac{m^2_2}{3}\leq m^2_3<m^2_2 
\label{eq:conditions} 
\eea 
the squared masses $M_1^2$ and $M_2^2$ are real positive, so that the potential is a sum of two 
Yukawa terms: 
\bea 
\Phi&=&\frac{1}{r}( A_1\,e^{-M_1\,r}+A_2\,e^{-M_2\,r}) 
\eea 
Also, if one of the $A_{1,2}$ vanishes, the potential mimics a single Yukawa term of the FP theory.

The conditions (\ref{eq:conditions}), via (\ref{massval}) and the similar expressions for $m_2$,
$m_3$, turn into point-wise constraints on the derivative of $V$ on Minkowski space, which can be always
satisfied by appropriate choices of the two otherwise arbitrary functions $\U$ and $\E$.

By (\ref{eq:yuk}) and by inspection of the other gauge invariant potential in the metric
fluctuation, one finds that both potentials reduce to $1/2\Delta$ at short distance, confirming the
absence of vDVz discontinuity ~\cite{Rubakov}. The phenomenological advantages of having a weak
coupled theory compared with the Lorentz invariant case has been discussed in~\cite{usweak}.

 \section{Conclusions}
\label{conc}

In this this paper, by using the canonical Hamiltonian analysis, we studied the number of
propagating degrees of freedom in generic theories of massive gravity.  Such theories which lead to
large distance modifications of gravity, include nonderivative terms in the action, encoded in a
potential $V$ function of the metric components.  To build $V$ extra tensorial objects are
needed. We show that, in order to preserve at least the rotational symmetry SO(3) and lead to a
massive graviton, at least an extra 3-dimensional metric is needed. In the minimal approach such a
metric is non dynamical, and will lead to potentially observable preferred frame effects.

From the canonical analysis, it follows that 6 DoF propagate in general unless additional conditions
on $V$ are introduced.  In particular we have shown that 5 DoF propagate if and only if the
potential satisfies the Monge-Ampere equation (\ref{monge}) plus the additional differential
equation (\ref{pilo}). Such set of differential equations can be solved in terms of two arbitrary
functions which  parametrize the most general nonderivative gravity modification with 5 DoF at full
non perturbative level. Thus, by using the canonical analysis, the problem of finding the most
general theory of modified massive gravity with 5 DoF has been completely solved.  Let us summarize
the result in a nutshell: given two arbitrary functions $\U(\K^{ij})$ and $\E(\xi^i,\g^{ij})$ (of
the specific arguments) the deformation potential $V$ of any massive
gravity theory with  five propagating DoFs can be written as
\bea
  V&=&\U +N^{-1} \left(\E+\U_i\,\Q^i \right)\qquad {\rm where}\nonumber\\
 && N^i=N\,\xi^i+\Q^i\,,\qquad \Q^i(\xi^i,\g^{ij})\equiv-\,\U_{ij}^{-1}\,\E_j\,,\qquad
 \K^{ij}\equiv\g^{ij}-\xi^i\,\xi^j \, .\nonumber
\eea
In addition, $\E$ represents the energy (Hamiltonian) density of the system and thus it can be
chosen to be positive definite.

Besides its theoretical interest, this result is also relevant from a phenomenological point of
view, for a number of reasons.  First, because it uncovers a large class of massive gravity theories
that are ghost free on Minkowski space, whereas previously, the only known ghost free theory was the
Lorentz-invariant four parameter theory of~\cite{Gabadadze:2011,GF}, which is shown here to be a
very special case of our general construction.

Second, while Lorentz symmetry may be enforced, the price to be paid is the impossibility of using
perturbation theory in many physical important situations like inside our solar system. Instead, in
generic Lorentz-breaking theories this problem disappears.  It ought to be remarked that the Lorentz
symmetry we are discussing here only concerns the gravitational sector, and is not the same symmetry
that enters in the formulation of the Einstein's equivalence principle. As such, is not subject to
phenomenological constraints. Thus, we conclude that the viability of the theory is directly
connected with the need to have Lorentz-breaking in the gravitational sector, which should also be
testable at forthcoming gravitational wave experiments.

The concrete phenomenology of the new class of Lorentz breaking theories is also rather promising,
as argued in~\cite{usweak}. From a perturbative point of view, from the general expression of $V$,
we have found that remarkably some relations among the graviton masses exists and are important in
view of the phenomenological application. From a nonperturbative side, besides the absence of ghosts
in the spectrum, crucially important features are the possibility to trust the theory up to the
cutoff $\Lambda_2=(m M_{PL})^{1/2}\simeq (10^{-3} \,\text{mm})^{-1}$, and the absence of strong
nonlinearities (Vainshtein radius) around macroscopic sources.

It is crucial to study  the impact of the general construction
presented here on cosmology. We leave the matter for a future publication~\cite{uscosm}.

\begin{appendix}
\section{Canonical analysis made simple in the rank 3 case}

In the Monge-Ampere case, the canonical analysis is made very simple and transparent if the $\xi^i$
variables are used in place of the shifts $N^i$, through the change of variables
\be
N^i = N\xi^i+Q^i(\xi^i,\g^{ij})\,.
\ee

The Hamiltonian, as a function of the $(N,\xi)$ variables, is 
\be
H=N (\H_0+\xi^i\H_i) + Q^i\H_i+\V(N,N\xi^i+Q^i)\,.
\ee
The primary constraints are as usual $\Pi_A=\partial H/\partial (\dot N,\dot \xi^i)\approx 0$, and
one introduces four Lagrange multipliers $\l^A$ and a total hamiltonian $H_T=H+\lambda^A\Pi_A$.

The secondary constraints are now modified, but after using the
Monge-Ampere relations~(\ref{vc}) and (\ref{Qrel}), one has
the following simple expressions
\bea
\label{eq:newS0}
S_0&=&\H_0+\xi^i\;\H_i + \sqrt{\gamma}\; \U \approx 0  \, ;\\
\S_i&=&\H_i+ \sqrt{\gamma}\;\U_i \approx 0 \, .
\eea
In the simplified form the secondary constraints are expressed in
terms of the function $\U$, which depends only on $\xi$ and not on
$N$. Notice that   $\S_0=\S_0^{old}+\xi^i\;\S_i\approx S_0^{old}$.
The tertiary constraints are 
\bea
\T_0&=&\{\S_0,H\} \approx 0 \, ;\\
\T_i&=&\{\S_i,H\}+\V_{ij}\lambda^j \approx 0 \, .
\eea
 The Lagrange multiplier $\l^0$ drops out from the expressions, and this is due to the
fact that $\U$ on $\S_0$ does not depend on $N$.  Thus, $\T_0$ represents a genuine new constraint,
eliminating half of the sixth mode in phase space.

Finally, the time evolution of $\T_0$ is
\be
{\cal Q}=\{\T_0,H_T\}= \{\S_0,\S_0\}\;\l^0 + \text{($\l^0$-independent
  terms)} \approx 0 \, ;
\ee
and if one requires that $\{\S_0,\S_0\}=0$, we get  a further constraint which completes the
elimination of the sixth mode.  This condition, using (\ref{eq:newS0}) and the algebra of GR (see
appendix~\ref{comm}), leads straightforwardly to the additional condition (\ref{five}).

Thus, the implicit change of variables~(\ref{Qrel}) from $N^i$ to $\xi^i$ has the effect of reducing
the hessian null eigenvector to $(1,\vec 0)$, and it helps to make more transparent the generation of the
new constraints.

\section{Poisson Brackets}
\label{comm}

We calculate here explicitly the poisson brackets required for 
\be
\begin{split}
\Theta(x,y)=& \chi^A(x)  \chi^B(y) \, \{ \S_A(x), \, \S_B(y)
\} \\
=&  \chi^0(x)  \chi^0(y) \, \{ \S_0(x), \, \S_0(y)
\} \\
&+  \chi^0(x)  \chi^j(y) \, \{ \S_0(x), \, \S_j(y) \}  + \chi^i(x)  \chi^0(y) \, \{ \S_i(x), \, \S_0(y)
\} \\
&+  \chi^i(x)  \chi^j(y) \, \{ \S_i(x), \, \S_j(y) \}  .
\end{split}
\ee
Using the following standard expression $\H$ and $\H_i$
for the Hamiltonian and
momentum constraints in GR 
\be
\begin{split}
&\H(x) = \frac{1}{\sqrt{\gamma}} \left[2 \Pi_{mn}  \Pi^{mn}- (\gamma^{ij}
  \Pi_{ij})^2 \right]-\sqrt{\gamma} R^{(3)}  \; ;  \\
&\H_i(x) = -2 \gamma_{im} D_k \Pi^{mk} \, ;
\end{split} 
\ee
we have that
\be
\begin{split}
\{ \S_0(x), \, \S_0(y) \}&= \H^i(x) \de_i^{(x)}
  \delta^{(3)}(x-y) -  \H^i(y) \de_i^{(y)} \delta^{(3)}(x-y) \\
&+
 \left[ \F_{mn}(x) \V_{\gamma N}^{mn}(y)-\F_{mn}(y) {\cal
     V}_{N}^{mn}(x) \right] \delta^{(3)}(x-y)  \; ;
\end{split}
\ee
\be
\begin{split}
\{ \S_0(x), \, \S_j(y) \}&= \H(y) \de_j^{(x)} \dd  
+ \dd \, \F_{mn}(x)  \V_j^{mn}(y) \\
&+    \V_N^{ab}(x) \, Q_{abj}(y)  \dd -2 \,
\V_N^{ab}(x) \, 
\gamma_{j(a}(y) \de_{b)}^{(y)} \dd \, .
\end{split}
\ee
\be
\begin{split}
\{ \S_i(x), \, \S_0(y) \}&= 
 -\H(x) \de_i^{(y)} \dd  
- \dd \, \F_{mn}(y)  \V_i^{mn}(x) \\
&-    \V_N^{ab}(y) \, Q_{abi}(x)  \dd +2 \,
\V_N^{ab}(y) \, 
\gamma_{i(a}(x) \de_{b)}^{(x)} \dd \, .
\end{split}
\ee
\be
\begin{split}
 \{ \S_i(x), \, \S_j(y) \}&= \H_j(x) \de_i^{(x)} \dd  - \H_i(y) \de_j^{(x)} \dd\\
& -\dd \, \V_j^{ab}(y) \, Q_{abi}(x) + 2  \V_j^{ab}(y) \, 
\gamma_{i(a}(x) \de_{b)}^{(x)} \dd \\
& + \dd  \V_i^{ab}(x) \, Q_{abj}(y) -2   \V_j^{ab}(y) \, 
\gamma_{j(a}(y) \de_{b)}^{(y)} \dd \, .
\end{split}
\ee
where
\be
\begin{split}
&\F_{mn}(x) = \gamma^{-1/2} \left( 2 \Pi_{mn}(x) - \Pi^a_a(x)
  \gamma_{mn}(x) \right) \, ; \\
&Q_{abm}(x) = \de^{(x)}_m \gamma_{ab}(x) -2 \de_{(a}^{(x)}
\gamma_{b)m}(x) \, ;\\
& \V_N^{mn} = \frac{\de^2\V}{\de N \de \gamma_{mn} } \, , \qquad
\V_i^{mn} = \frac{\de^2 \V}{\de N^i  \de \gamma_{mn} }
\, .
\end{split}
\ee
We have also used the following relation for the Lie derivative $\L$
of spatial metric with respect of a spatial vector $\epsilon$
\be
\left(\L_\epsilon \gamma \right)_{ab} =D_a \epsilon_b + D_a
\epsilon_a =\epsilon^m \de_m \gamma_{ab} + \gamma_{mb} \de_a
\epsilon^m + \gamma_{am} \de_b \epsilon^m \, .
\ee
Assembling the various pieces we arrive at the result 
\be
\Theta(x,y) = A^i(x,y) \, \de_i^{(y)}  \dd + B^i(x,y) \, \de_i^{(x)}  \dd  +C(x,y) \, \dd \, ;
\ee 
with
\be
\begin{split}
A^k(x,y) = & - \chi^0(x) \chi^0(y) \,  \H^k(y) - 2 \chi^0(x)
\chi^j(y)  \, \gamma_{ja}(y) \, \V_N^{ak}(x) 
-\chi^i(x) \chi^k(y)  \,\H_i(y) \\
&- 2 \chi^i(x) \chi^j(y)  \, \gamma_{ja}(y) \, \V_i^{ak}(y) - \chi^k(x)
\chi^0(y)  \,  \H(x) \, ;
\end{split}
\ee
\be
\begin{split}
B^k(x,y) = &  \chi^0(x) \chi^0(y) \,  \H^k(x) + \chi^k(y)
\chi^0(x)  \,  \H(y) + 2 \chi^0(y)
\chi^i(x)  \, \gamma_{ia}(x) \, \V_N^{ak}(y)\\ 
&+ 2\chi^k(x) \chi^j(y)  \,\H_j(x) 
+ \chi^i(x) \chi^j(y)  \, \gamma_{ia}(x) \, \V_j^{ak}(x)  \, ;
\end{split}
\ee
and
\be
\begin{split}
C(x,y) = & \chi^0(x) \chi^0(y )\left[ {\cal C}_{mn}(x) \V_N^{mn}(y)-  {\cal C}_{mn}(y) \V_N^{mn}(x) \right] \\
&+\chi^0(x) \chi^i(y )\left[ {\cal C}_{mn}(x) \V_i^{mn}(y)+  {\cal
    V}_N^{ab}(x)  \, Q_{abi}(y) \right] \\
&- \chi^i(x) \chi^0(y )\left[ {\cal C}_{mn}(y) \V_i^{mn}(x)+  {\cal
    V}_N^{ab}(y)  \, Q_{abi}(x) \right]\\
&+ \chi^i(x) \chi^j(y )\left[ \V_i^{ab}(x) \, Q_{abj}(y) - {\cal
    V}_j^{ab}(y)  \, Q_{abi}(x) \right] \, .  
\end{split}
\ee
Thus
\be
\begin{split}
&I(x) = \int d^3y \, z(y) \, \Theta(x,y) =D^i(x,x) \, \de_i z(x) + z(x)
\, \de^{(y)}_i D^i(x,y)_{| y=x} \, ; \\
& D^i(x,y) = A^i(x,y) - B^i(x,y) \, .
\end{split}
\ee
The same result can be also obtained by using $ \de_j^{(x)} \dd= -
\de_j^{(y)} \dd$.

\no
We notice that
\be
\begin{split}
&C(x,y) = -  C(y,x) \Rightarrow C(x,x) =0 \, ; \\
& B^i(x,y) =- A^i(y,x)  \Rightarrow D^i(x,y) = A^i(x,y) + A^i(y,x)
\Rightarrow D^i(x,y) = D^i(y,x) \, . 
\end{split}
\ee
From the symmetry of $D^i$, it follows that $\de^{(y)}_i
D^i(x,y)_{|y=x}= 1/2 \, \de_i^{(x)} D^i(x,x)$ and as  a result
\be
I(x) = \int d^3y \, z(y) \, \Theta(x,y) =
\frac{1}{2 \, z(x)}  \de_i \left[ z(x)^2 \, D^i(x,x) \right ] \, .
\ee
The integral $I(x)$ vanishes for all $z$ if $D^i(x,x)=2 A^i(x,x)
=0$, thus the above condition requires
\be
  2\chi^A \chi^m \V_A^{in} \gamma_{nm} 
 - {\chi^0}^2  \V_j \gamma^{ij} -  \chi^i \chi^A    \V_A
 =0  \, ,
\ee
which is precisely (\ref{add}).

\section{Quaternary constraint}
\label{cantime}
In the case $r=\text{Rank}(\V_{AB}) =3$, the only
tertiary $ \T$ constraint reads
\be
 \T \equiv \frac{d }{dt} (\chi^A \S_A)=\chi^A \, \left[ \left \{ \S_A, \, \H \right \} -
\lambda^B \, \V_{AB} \right]
\approx 0 \, .
\ee
We have to impose that also $\T$ is conserved 
\be
\frac{d  \T}{dt} =
\left \{
 \T, \, H \right \} + \int d^3 y \, \left \{ 
\T(x), \,  \lambda^A(y) \Pi_A(y) 
\right \} \, .
\ee
The first two terms do not depend on $z$; using Jacobi identity, the last term can be written as
\be
\begin{split}
 \left \{   \T(x), \,  \lambda^A(y) \, \Pi_A(y) \right \}
 &= \lambda^A\left [ \frac{\de \chi^B}{\de N^A} \, \{ \S_B, \, H \} + \{ \V_{AB},H \} \chi^B \right]\\
&+ \int d^3y \, \lambda^A(y) \, \chi(x)^B \left \{ \S_A(y), \, 
\S_B(x) \right \}  \, .
\end{split}
\ee
Now using that $\lambda^A = z \, \chi^A + \bar \lambda^A$, the
properties of $\chi^A$ and the previous constraints we have that
\be
\begin{split}
&\frac{\de \chi^B}{\de N^A} \, \{ \S_B, \, H \} = - \frac{\de
  \chi^B}{\de N^A}  \left (  \V_{BC} \, \bar \lambda^A \, \bar
  \lambda^C  \right) \, ; \\
& \lambda^A \,  \{ \V_{AB}, \, H \} \chi^B  = \bar  \lambda^A  \,    \{
\V_{AB}, \, H \} \chi^B \, .\\
\end{split}
\ee
Thus
\be
\begin{split}
 &\int d^3y \, \left \{  
\tilde \T(x), \,  \lambda^A(y) \, \Pi_A(y) 
\right \}
 = 
-  \frac{\de
   \chi^B}{\de N^A}   \, \V_{BC} \, \bar \lambda^A \, \bar
  \lambda^C +  \bar  \lambda^A  \,    \{ \V_{AB}, \, H \} \chi^B \\
& + \int d^3y \, \bar \lambda^A(y) \, \chi^B(x) \, \{ S_A(y), \,
\S_B(x) \}  - \int d^3y \,  z(y) \, \Theta(x,y) \, .\\
\end{split}
\ee
As a result the quaternary constraint does not depend on $z$ if (\ref{add}) holds, and it gives a
genuine new constraint.

\section{General solution of the Monge-Ampere Equation}
\label{int}
The solution of the Monge-Ampere equation $\tilde \V$ is given
implicitly in terms of the two functions $\U$, $\L$ of $\xi$ and
$\gamma$ by 
\be
\frac{\de \tilde \V}{\de N} = \U - \frac{\de \U}{\de \xi^i} \xi^i \,
,\qquad \frac{\de \tilde \V}{\de N^i}= \frac{\de  \U}{\de \xi^i} \, .
\label{def}
\ee
The relation between $N^i$ and $\xi$ reads
\bea
\label{Qrel1}
 N^i&=&N\; \xi^i+\Q^i\\
\Q^i(\xi^i,\gamma^{ij})&=&-\left(\U_{ij}\right)^{-1}\E_j \, .
\eea 
Integrating the first of (\ref{def})  with respect to $N$ we can exploit
the relation that follows from (\ref{Qrel1})
\be
- \frac{\de  \U}{\de \xi^i} \xi^i = N \frac{\de  \U}{\de N}+
\frac{\de Q^i}{\de N } \frac{\de  \U}{\de \xi^i} \, .   
\ee
Thus
\be
\begin{split}
\tilde \V &= \int dN \left ( \U  - \frac{\de  \U}{\de \xi^i}
  \xi^i \right)= \int dN \left ( \U  + N \frac{\de  \U}{\de N}+
\frac{\de Q^i}{\de N } \frac{\de  \U}{\de \xi^i} \right) =\\
&Q^i
\frac{\de \U}{\de \xi^i} + \int dN \left
[\frac{\de} {\de N} ( N \U ) - Q^i  \frac{\de^2  \U}{\de \xi^i 
\de N} \right] \, .
\end{split}
\ee
Now, from the definition of $Q^i$ we have that 
\be
Q^i  \frac{\de^2  \U}{\de \xi^i 
\de N} = Q^i  \frac{\de^2  \U}{\de \xi^i 
\de \xi^j} \frac{\de \xi^j}{\de N} = -\frac{\de \E}{\de \xi^i}
\frac{\de \xi^i}{\de N} =-\frac{\de \E}{\de N} \, .
\ee
As a result
\be
\tilde \V = N \U + \E + Q^i
\frac{\de \U}{\de \xi^i} + \C \, ;
\ee
where $\C$ is function of $N^i$ and $\gamma_{ij}$. 
One can check that  the above expression for $\tilde \V$ 
satisfies  the last of (\ref{def}) when $\C=0$.
Thus we get the final expression: $\tilde \V=  N \U + \E + Q^i \U_i$.

\end{appendix}


\begin{thebibliography}{99}
\bibitem{Fierz:1939ix}
  M.~Fierz and W.~Pauli,
  Proc.\ Roy.\ Soc.\ Lond.\  A {\bf 173}, 211 (1939).


\bibitem{BD}
  D.~G.~Boulware and S.~Deser,
  Phys.\ Lett.\  B {\bf 40}, 227 (1972).


\bibitem{uscan} 
  D.~Comelli, M.~Crisostomi, F.~Nesti and L.~Pilo, 
 Phys.\ Rev.\  D {\bf 86} (2012) 101502.  [arXiv:1204.1027 [hep-th]].

\bibitem{Soloviev:2013mia} 
  V.~O.~Soloviev and M.~V.~Tchichikina,
  arXiv:1302.5096 [hep-th].


\bibitem{Gabadadze:2011}  
C.~de Rham, G.~Gabadadze, A.~J.~Tolley,
  Phys.\ Rev.\ Lett.\  {\bf 106}, 231101 (2011).
  [arXiv:1011.1232 [hep-th]].

\bibitem{GF}
S.~F.~Hassan and R.~A.~Rosen,
  Phys.\ Rev.\ Lett.\  {\bf 108}, 041101 (2012)\\
  S.~F.~Hassan, R.~A.~Rosen and A.~Schmidt-May,
  JHEP {\bf 1202}, 026 (2012)
  [arXiv:1109.3230 [hep-th]].
.
S.~F.~Hassan and R.~A.~Rosen,
  arXiv:1111.2070 [hep-th].


\bibitem{DGT} 
  C.~de Rham, G.~Gabadadze and A.~Tolley,
  arXiv:1107.3820 [hep-th].


\bibitem{Zumino:1970tu}
  B.~Zumino,
  In *Brandeis Univ. 1970, Lectures On Elementary Particles And Quan tum Field Theory, Vol. 2*, Cambridge, Mass. 1970, 437-500.


\bibitem{DIS}
H.~van Dam and M.~J.~G.~Veltman,
\emph{Nucl.\ Phys.\ }  B {\bf 22} (1970) 397;
Y.~Iwasaki,
\emph{Phys.\ Rev.\ }  D {\bf 2} (1970) 2255;\\
 V.I.Zakharov, \emph{JETP Lett.} {\bf 12} (1971) 198.

\bibitem{Vainshtein}
  A.~I.~Vainshtein,
Phys.\ Lett.\  B {\bf 39}, 393 (1972);\\

\bibitem{vain2} 
  E.~Babichev, C.~Deffayet and R.~Ziour, 
  \emph{Phys.\ Rev.\ Lett.}\  {\bf 103}, 201102 (2009).\\ 
  N.~Kaloper, A.~Padilla and N.~Tanahashi, 
  \emph{JHEP} {\bf 1110}, 148 (2011).\\ 
  G.~Chkareuli and D.~Pirtskhalava,
  Phys.\ Lett.\ B {\bf 713}, 99 (2012)
  [arXiv:1105.1783 [hep-th]].

\bibitem{padilla} 
  C.~Burrage, N.~Kaloper and A.~Padilla,
  arXiv:1211.6001 [hep-th].

\bibitem{Deser}
  S.~Deser and A.~Waldron,
  arXiv:1212.5835 [hep-th].


\bibitem{Rubakov}
V.~A.~Rubakov,
arXiv:hep-th/0407104.





\bibitem{dub}
  S.~L.~Dubovsky,
JHEP {\bf 0410}, 076 (2004);\\
V.~A.~Rubakov and P.~G.~Tinyakov,
  Phys.\ Usp.\  {\bf 51}, 759 (2008).

\bibitem{diego} 
 D.~Blas, D.~Comelli, F.~Nesti, L. Pilo, 
 \emph{Phys.\ Rev.}\  {\bf D80}, 044025 (2009),  
 arXiv:0905.1699 [hep-th]. 

\bibitem{PRLus}
  Z.~Berezhiani, D.~Comelli, F.~Nesti and L.~Pilo,
  Phys.\ Rev.\ Lett.\  {\bf 99}, 131101 (2007)

\bibitem{gaba}   
  G.~Gabadadze and L.~Grisa,
  Phys.\ Lett.\  B {\bf 617} (2005) 124;
L.~Grisa,
JHEP {\bf 0811} (2008) 023.

\bibitem{usweak}
  D.~Comelli, F.~Nesti and L.~Pilo,
arXiv:1302.447 [hep-th].


\bibitem{HGS}
  N.~Arkani-Hamed, H.~Georgi and M.~D.~Schwartz,
  Annals Phys.\  {\bf 305}, 96 (2003).

\bibitem{DAM1}
 C.J.~Isham, A.~Salam and J.A.~Strathdee,
  \emph{Phys.\ Rev.}  D {\bf 3}, 867 (1971).
A.~Salam and J.~A.~Strathdee,
   \emph{Phys.\ Rev.}  D {\bf 16} (1977) {2668};\\
C.~Aragone and J.~Chela-Flores,
  \emph{Nuovo Com.} {\bf A10} (1972) {818};\\
%
  T.~Damour and I.~I.~Kogan,
  Phys.\ Rev.\  D {\bf 66} (2002) 104024.
[arXiv:hep-th/0206042].

\bibitem{will} 
  C.~M.~Will,
  Living Rev.\ Rel.\  {\bf 9}, 3 (2006)
  [gr-qc/0510072].


\bibitem{myproc} 
 L.~Pilo,
  PoS EPS {\bf -HEP2011}, 076 (2011).

\bibitem{spher} 
  D.~Comelli, M.~Crisostomi, F.~Nesti and L.~Pilo,
  Phys.\ Rev.\ D {\bf 85}, 024044 (2012)
  [arXiv:1110.4967 [hep-th]].
\bibitem{energy}
  D.~Comelli, M.~Crisostomi, F.~Nesti and L.~Pilo,
  Phys.\ Rev.\ D {\bf 84}, 104026 (2011)
  [arXiv:1105.3010 [hep-th]].

\bibitem{Volkov} 
  M.~S.~Volkov,
  Phys.\ Rev.\ D {\bf 85}, 124043 (2012)
  [arXiv:1202.6682 [hep-th]].


\bibitem{spher1} 
  Z.~Berezhiani, D.~Comelli, F.~Nesti and L.~Pilo,
  JHEP {\bf 0807}, 130 (2008)
  [arXiv:0803.1687 [hep-th]].

\bibitem{ADM} 
  R.~L.~Arnowitt, S.~Deser and C.~W.~Misner,
  gr-qc/0405109.


 







\bibitem{Henneaux:2010vx} 
  M.~Henneaux, A.~Kleinschmidt and G.~Lucena Gomez,
  arXiv:1004.3769 [hep-th].


\bibitem{MongeAmpere}
D. Gilbarg and N.S. Trudinger, ``Elliptic Partial Differential
Equations of Second Order''. Berlin: Springer-Verlag, 1983
A.V. Pogorelov (2001), ``Monge-Amp\`ere equation'', in Hazewinkel,
Michiel, Encyclopedia of Mathematics, Springer.

\bibitem{fai}
D.B.~Fairlie, A.N.~Leznov,
Journal of Geometry and Physics  {\bf 16} 385 (1995).  hep-th/9403134.


 
\bibitem{Berezhiani:2009kv}  
  Z.~Berezhiani, F.~Nesti, L.~Pilo and N.~Rossi, 
  JHEP {\bf 0907}, 083 (2009) 
  [arXiv:0902.0144 [hep-th]]. 


\bibitem{uscosm} 
 D.~Comelli, F.~Nesti and L.~Pilo,
  JCAP {\bf 1405}, 036 (2014)
  [arXiv:1307.8329 [hep-th]].

\end{thebibliography}
\end{document}